\documentclass[
reprint,
superscriptaddress,
showpacs,
aps,
]{revtex4-1}

\usepackage{amsmath}
\usepackage{amssymb}%
\usepackage{natbib}
\usepackage{graphicx}
\usepackage{dcolumn}
\usepackage{bm}
\usepackage{ulem,color}

\usepackage{ulem} 

\newcommand{\mrm}{\mathrm}
\newcommand{\mcl}{\mathcal}
\newcommand{\mbb}{\mathbb}

\newcommand{\lt}{\left}
\newcommand{\rt}{\right}
\newcommand{\lag}{\langle}
\newcommand{\rag}{\rangle}

\newcommand{\hspc}{\hspace{1em}}
\newcommand{\htab}{\hspace{2em}}
\newcommand{\intra}{\mrm{intra}}
\newcommand{\inter}{\mrm{inter}}
\newcommand{\im}{\mrm{i}}

\begin{document}

\title{Cluster-Based Haldane States in Spin-1/2 Cluster Chains}
 
\author{Takanori Sugimoto}
\email{sugimoto.takanori@rs.tus.ac.jp}
\affiliation{Department of Applied Physics, Tokyo University of Science, Katsushika, Tokyo 125-8585, Japan}
\author{Katsuhiro Morita}
\affiliation{Department of Applied Physics, Tokyo University of Science, Katsushika, Tokyo 125-8585, Japan}
\author{Takami Tohyama}
\affiliation{Department of Applied Physics, Tokyo University of Science, Katsushika, Tokyo 125-8585, Japan}

\date{\today}

\begin{abstract}
The Haldane state is a typical quantum and topological state of matter, which exhibits an edge state corresponding to symmetry-protected topological order in a one-dimensional integer spin chain. This edge state can be utilized for a processing unit of quantum computation. Its realization, however, has difficulties with synthesis of integer spin compounds. In contrast, one-half spin systems are more designable due to recent progress on intended synthesis of organic materials, quantum dots, and optical lattices. Here we propose a concept to design the Haldane state with one-half spins by making use of a chain composed of one-half spin clusters. If the clusters contains two spins, the ground state of a chain corresponds to the Affleck--Kennedy--Lieb--Tasaki state.
In the case of an odd number of spins in the clusters, we propose a concrete procedure to construct a field-induced Haldane state. We illustrate the procedure with a 5-spin cluster chain.
\end{abstract}

\pacs{}
\maketitle

The concept of topology has given us clear perspectives on the state of matter annoyed by quantum and/or thermal fluctuations.
In condensed matter, the topology provides robust physical properties against disturbance of circumstance.
For instance, topologically-protected quantum computing has been proposed in a topological superconductor~\cite{Read00,Kitaev01,Ivanov01,Mazza13}, and a flat band in topological matters is expected to heighten the critical temperatures of fractional quantum Hall state~\cite{Tang11,Sun11,Neupert11} and surface superconductor~\cite{Kopnin11}. 
Among many topological matters, one of the simplest models is the Haldane state in a quantum spin chain in which an integer spin is antiferromagnetically coupled to the nearest-neighbor spins.
This state has been firstly proposed by F. D. M. Haldane in 1983 with a famous conjecture~\cite{Haldane83}: an antiferromagnetic spin chain of integer spins ($S=1,2,3,\cdots$)~\cite{Note0} has a finite excitation gap above the ground state.
As a characteristic of the $S=1$ Haldane state, Affleck--Kennedy--Lieb--Tasaki (AKLT) have shown a hidden topological order, so-called {\it string} order, in a clear schematic of the rigorous ground state~\cite{Affleck88,Nijs89,Tasaki91,Kennedy92}.
In this theory, an $S=1$ spin is decomposed into two {\it virtual} $S=\frac{1}{2}$ spins.
As the ground state, the virtual $S=\frac{1}{2}$ spins of neighboring sites are anti-symmetrized, and the virtual spins at each site are symmetrized as a real $S=1$ spin.
Hence, the ground state retains a strong entanglement between neighboring sites, where a finite string order appears.
Furthermore, F. Pollmann {\it et al.} have shown a qualitative diffrenece between $S=\>$odd and even Haldane chains~\cite{Pollmann12}, which is so-called symmetry-protected topological order.
Recently, G. K. Brennen and A. Miyake have proposed the concept of holographic quantum computing using the symmetry-protected topological order in the Haldane state~\cite{Miyake08}.
There are, however, few compounds possessing the $S=\>$odd spins in one dimension~\cite{Miyake08,Gross07,Else12}.
Here we propose a concept to design the $S=1$ Haldane chain with one-half spin clusters, which will be designable in organic materials~\cite{Ishiguro98}, quantum dots~\cite{Kane98,Loss98}, and optical lattices~\cite{Cirac95,Greiner02}. 
Moreover our concept enables us to directly observe the entangled quantum state of one-half spins in the Haldane state, which is applicable to the quantum computing at the dawn of designable quantum states based on one-half spins.

In the present study, we call a model given by the following Hamiltonian a spin cluster chain (SCC),
\begin{equation}
\mcl{H}_0 = \sum_{j=1}^L \mcl{H}_{\mrm{intra}}^{(j)} + \sum_{j=1}^{L-1} \mcl{H}_{\mrm{inter}}^{(j)}
\end{equation}
with
\begin{align}
&\mcl{H}_{\mrm{intra}}^{(j)} = \sum_{\lag i,i^\prime\rag \in B_{\mrm{intra}}} J_{\lag i,i^\prime\rag}\bm{S}_{i,j}\cdot\bm{S}_{i^\prime,j}, \\
&\mcl{H}_{\mrm{inter}}^{(j)} = \sum_{\lag\lag i,i^\prime\rag\rag \in B_{\mrm{inter}}} J_{\lag\lag i,i^\prime\rag\rag}^\prime \bm{S}_{i,j}\cdot\bm{S}_{i^\prime,j+1},
\end{align}
where $L$ is the number of clusters, and $B_{\mrm{intra}}=\{(i,i^\prime)\,|\,i,i^\prime\in \mbb{N},1\leq i<i^\prime\leq N\}$ and $B_{\mrm{inter}}=\{(i,i^\prime)\,|\,i,i^\prime\in \mbb{N},1\leq i\leq N, 1\leq i^\prime\leq N\}$, with $N$ being the number of spins in a cluster,  are a set of intra cluster bonds and a set of inter cluster bonds, respectively.
For simplicity, we consider only the Heisenberg interactions as interactions between one-half spins.
As a condition of cluster chain, we assume that the intra cluster interaction $J_\intra=\sqrt{\sum_{\lag i,i^\prime\rag \in B_{\mrm{intra}}} (J_{\lag i,i^\prime\rag})^{2}}$ is much greater than the inter cluster interaction $J_\inter=\sqrt{\sum_{\lag\lag i,i^\prime\rag\rag \in B_{\mrm{inter}}} (J_{\lag\lag i,i^\prime\rag\rag}^\prime)^2}$, i.e., $J_\intra\gg J_\inter$.
Figure 1 shows the ground state of this model schematically, where a transparent ball including arrows represents a spin cluster, which is connected by a weak inter cluster interaction (a transparent bond between balls).
Temperature dependence in this system is divided into three stages by two characteristic temperatures $T_1 < T_2$:
\newpage 
\begin{itemize}
\item High-temperature region ($T>T_2\sim J_\intra$) \\
Every spin behaves independently such like the Pauli paramagnetism. In Fig.~1, the transparent balls and bonds melt, and every spin (arrow) rotates freely.
\item Middle-temperature region ($T_1<T<T_2$)\\
The magnetic behavior in this region is understood by an isolated cluster of spins. Here the inter cluster interactions are effectively ignored by thermal fluctuation. In Fig.~1, spins included in a ball interact with each other, although the spins do not affect spins belonging to the next balls.
\item Low-temperature region ($T<T_1\sim J_\inter$)\\
A part of spin degrees of freedom is frozen, so that only the ground/low-energy states in a cluster contribute to many body interaction due to the inter cluster interaction. In Fig.~1, every ball is frozen, and thus spins in a ball do not move independently. However balls can rotate independently while the inter cluster interaction works on them. At zero temperature, the balls definitely stop due to the inter cluster interaction as expected in the ground state.
\end{itemize}

\begin{figure*}
\centering
\includegraphics[width=0.9\textwidth]{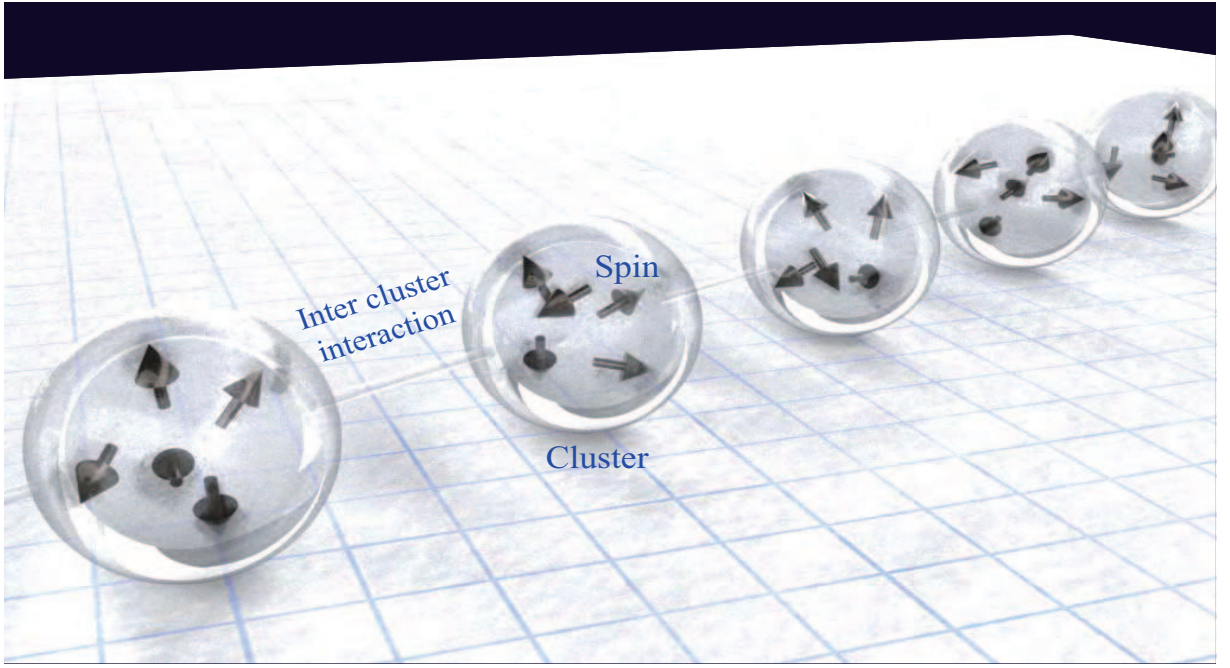}
\caption{Schematic ground state of a spin cluster chain (SCC). One-half spins ($S=\frac{1}{2}$) is represented by arrows. We assume that spins in a transparent ball (cluster) strongly interact with each other, and that the spins weakly interact with spins in the next balls. The detailed behaviors in three temperature regions are explained in the main text.}
\label{fig1}
\end{figure*}

Recent experimental study supported by our theoretical analysis has shown that a naturally occurring mineral Fedotovite K$_2$Cu$_3$O(SO$_4$)$_3$ is a SCC whose ground state is the Haldane state~\cite{Vergasova88,Starova91,Fujihala18}.
In this material, six $S=\frac{1}{2}$ spins form an edge-shared tetrahedral cluster, and the clusters stand in line.
The ground state in a cluster has been determined by the intra cluster interactions, which are estimated by the magnetization curve and the magnetic susceptibility in the middle temperature region ($T_1<T<T_2$).
Since the estimated interaction parameters have indicated that the ground state of Fedotovite is a triplet, the spin degree of freedom corresponding to an $S=1$ spin effectively survives in each cluster at low temperatures.
We thus expected the emergence of the Haldane state caused by a weak inter cluster interaction.
In fact, an inelastic neutron scattering experiment and magnetization measurement below $T_1\sim 4$~K have confirmed the existence of the first excitation gap originating from the inter cluster interaction.

What is the condition of the Haldane state in SCCs?
To clarify the condition, we start from an $S=\frac{1}{2}$ spin ladder, which exhibits the Haldane state as the ground state, if the rung interaction is ferromagnetic~\cite{Masuda06,Vekua06}.
Figure 2(a) shows the spin ladder, which corresponds to a 2-spin cluster chain whose cluster is a rung, if the cluster condition $|J^\prime|\ll |J|$ is satisfied.
When the intra cluster interaction is ferromagnetic $J<0$, the ground state is a triplet [Fig.~2(b)], and the low-energy physics is well described by an effective $S=1$ spin. 
Moreover, the inter cluster interaction is rewritten by a Heisenberg interaction between neighboring effective spins.
Hence, the effective model at low temperatures is a Heisenberg chain of effective $S=1$ spins, i.e., the $S=1$ Haldane chain.
Figure 2(c) shows the schematic ground state of the spin ladder for $J^\prime>0$.
In the ground state, the spin configuration of a rung is symmetric because of the triplet states, and that of neighboring rungs is anti-symmetric due to the antiferromagnetic inter cluster interaction.
Consequently, this ground state corresponds to the AKLT state, where the real $S=1$ spin is regarded as an effective $S=1$ spin of a rung.

\begin{figure}
\centering
\includegraphics[width=0.45\textwidth]{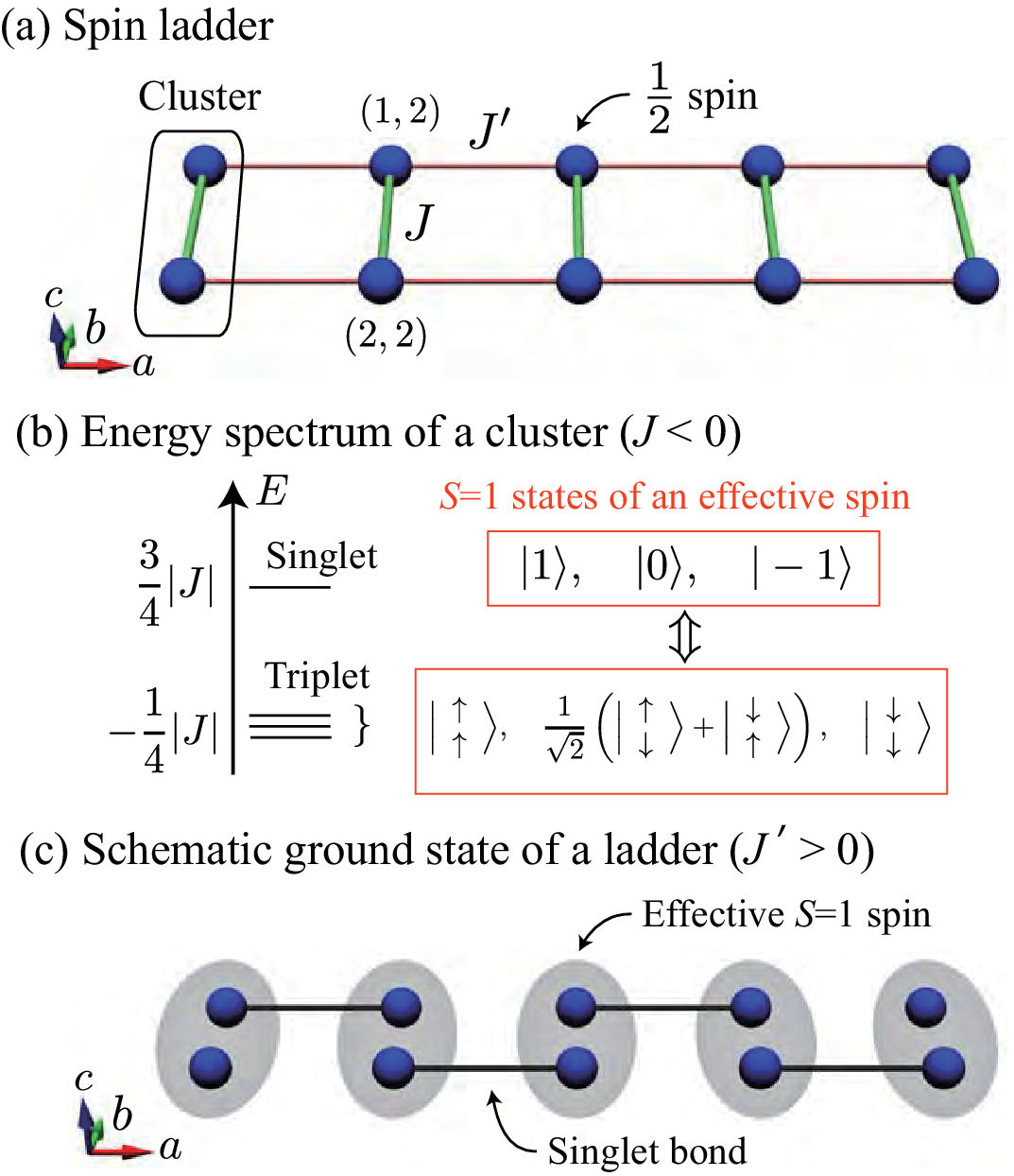}
\caption{(a) Spin ladder as a two-spin cluster chain (an $N=2$ SCC). The chain is elongated in the $a$ axis. Blue balls and colored segments denote the spin sites and interaction bonds, respectively. The interaction $J$ ($J^\prime$) corresponds to the intra (inter) cluster interaction. $(i,j)$ denotes the spin site index, e.g., $(1,2)$ means the 1st site in the 2nd cluster. (b) Energy spectrum of a cluster (rung) for $J<0$, where the ground state is a triplet. If the intra cluster interaction is much smaller than the inter cluster interaction $|J^\prime|\ll|J|$, only the triplet states of a cluster contributes to the low-energy physics as an effective $S=1$ spin. (c) Schematic ground state of a SCC for $J^\prime>0$. The triplet state of an effective $S=1$ spin is symmetric, and two spins of neighboring rungs are connected with the singlet (anti-symmetric) configuration. Thus, this ground state corresponds to the AKLT state.} 
\label{fig2}
\end{figure}

These examples intuitively give us the following conditions to realize the Haldane state in a SCC~\cite{Note1}.
\begin{enumerate}
\item[I.] The ground state of a cluster is triplet.
\item[II.] The inter cluster interaction is weak enough as compared with the first excitation gap in a cluster.
\item[III.] The effective inter cluster interaction between clusters' ground states is antiferromagnetic and planar-like (XY or Heisenberg type).
\end{enumerate}
We call the ground state satisfying these conditions {\it cluster-based} Haldane state (CBHS), where the string order defined by the effective spins should be observed.
In the CBHS, the string and spin correlations are defined by,
\begin{align}
C_{\mrm{str}}(r)=\lt\lag \tilde{S}_{j}^z\exp\lt[\im \pi \sum_{k=j+1}^{j+r-1} \tilde{S}_k^z \rt] \tilde{S}_{j+r}^z \rt\rag, \ C_{\mrm{spn}}(r)=\lt\lag \tilde{S}_j^z\tilde{S}_{j+r}^z \rt\rag,
\end{align}
where $\tilde{S}_j^z$ is the $z$ component of $S=1$ effective spin operator of $j$th cluster.
For the numerical calculation in this study, we set $j=\frac{L}{2}-\lt\lfloor\frac{r}{2}\rt\rfloor$, where the floor function $\lfloor x\rfloor$ represents the integer part of $x$. 
In the following, we illustrate the CBHS with two cases, where the number of spins $N$ in a cluster is either even or odd.

{\it First example: a 4-spin cluster chain (an $N=4$ SCC).} --- 
As explained above, the spin cluster in Fedotovite is edge-shared tetrahedra, where six $S=\frac{1}{2}$ spins form a cluster~\cite{Fujihala18}.
In the spin ladder model, the cluster contains two $S=\frac{1}{2}$ spins.
Both the cases correspond to the $N=\>$even condition, particularly $N=2$ (mod. 4).
Additionally, in Ref.~\cite{Fujihala18}, we have examined the CBHS in SCCs where the intra cluster interactions have the same symmetry as the cluster of Fedotovite.
In the SCCs, the conditions for the CBHS are satisfied when the number of spins in a cluster is $N=2$ (mod. 4).
Is the condition of $N=2$ (mod. 4) necessary to exhibit the CBHS?
The answer is no.
As an example of $N=0$ (mod. 4), we show the CBHS in a 4-spin cluster chain given by the following Hamiltonian,
\begin{align}
\mcl{H}_{\mrm{intra}}^{(j)} = &\frac{J_\intra}{2} \Big\{\lt[\mrm{Hs}^{(j)}_{(1,2)}-\mrm{Hs}^{(j)}_{(3,4)}\rt] + \cos\theta\lt[\mrm{Hs}^{(j)}_{(1,3)}+\mrm{Hs}^{(j)}_{(2,4)}\rt] \notag \\
&+\sin\theta\lt[\mrm{Hs}^{(j)}_{(1,4)}+\mrm{Hs}^{(j)}_{(2,3)}\rt]\Big\}
\end{align}
and 
\begin{equation}
\mcl{H}_{\mrm{inter}}^{(j)} = \frac{J_\inter}{\sqrt{2}} \lt[\cos\phi({\mrm{Hs}^\prime}^{(j)}_{(2,1)}+{\mrm{Hs}^\prime}^{(j)}_{(4,3)})+\sin\phi({\mrm{Hs}^\prime}^{(j)}_{(2,3)}+{\mrm{Hs}^\prime}^{(j)}_{(4,1)})\rt],
\end{equation}
where the intra (inter) cluster Heisenberg interaction $\mrm{Hs}_{(i,i^\prime)}^{(j)}=\bm{S}_{i,j}\cdot\bm{S}_{i^\prime,j}$ (${\mrm{Hs}^\prime}_{(i,i^\prime)}^{(j)}=\bm{S}_{i,j}\cdot\bm{S}_{i^\prime,j+1}$) and an angle $\theta$ ($\phi$) controls the ratio of intra (inter) cluster interactions.
This model is shown in Fig.~3(a).

\begin{figure*}
\centering
\includegraphics[width=0.9\textwidth]{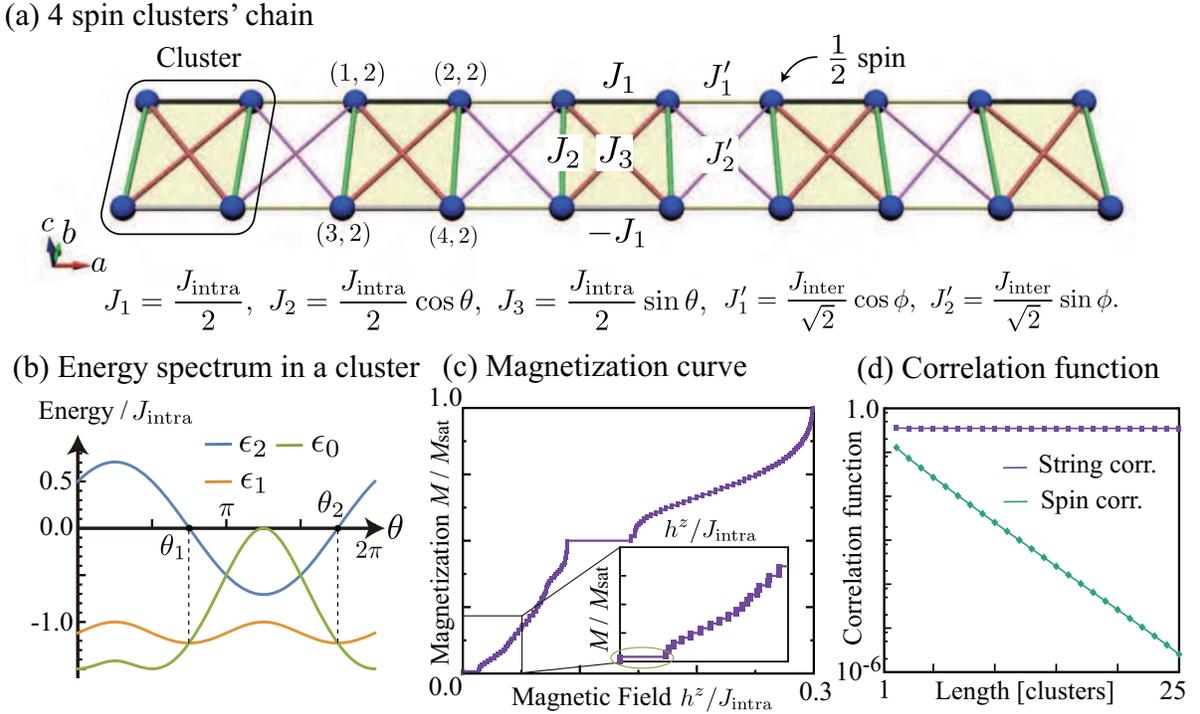}
\caption{(a) Four-spin cluster chain (an $N=4$ SCC). Notations are the same as Fig.~2(a). (b) Energy of a cluster as a function of $\theta$ that controls $J_i$ with fixed $\phi=\pi/4$. The lowest energy level of total spin ($S=$0, 1, and 2) is denoted by $\epsilon_S$. The energy levels of singlet and triplet states cross at $\theta_1=\frac{3}{4}\pi$ and $\theta_2=\frac{7}{4}\pi$. (c) Magnetization curve normalized by saturated magnetization $M_{\mrm{sat}}$ and (d) absolute value of string and spin correlation functions for $J_\inter/J_\intra=0.2$, $\theta=\frac{5}{4}\pi$, and $\phi=\frac{\pi}{4}$ in a 60-cluster chain ($L=60$) with open boundary condition. The inset of (c) is an enlarged view of the low field region, where we find the magnetization plateau at $M=1$ denoted by an ellipse. The correlation functions in (d) are calculated at $M=1$.}
\label{fig3}
\end{figure*}

Figure 3(b) shows that the ground state of a cluster is triplet [condition (I)] for $\theta \in [\theta_1,\theta_2]$.
Since the intra cluster Hamiltonian is symmetric with respect to the inversion $\sigma_j: (\bm{S}_{1,j},\bm{S}_{2,j},\bm{S}_{3,j},\bm{S}_{4,j})\to(\bm{S}_{2,j},\bm{S}_{1,j},\bm{S}_{4,j},\bm{S}_{3,j}$), the triplet ground state $|t^\gamma\rag_j$ ($\gamma=\pm, 0$) satisfies $\sigma_j|t^\gamma\rag_j =\pm |t^\gamma\rag_j$, resulting in
\begin{align}
\lag t^\gamma|_j (\bm{S}_{1,j}+\bm{S}_{3,j}) |t^\gamma\rag_j &=\lag t^\gamma|_j \sigma_j^2 (\bm{S}_{1,j}+\bm{S}_{3,j}) \sigma_j^2|t^\gamma\rag_j \notag\\
&=\lag t^\gamma|_j (\bm{S}_{2,j}+\bm{S}_{4,j}) |t^\gamma\rag_j.
\end{align}
We thus obtain the $S=1$ effective spin operators, $\tilde{S}_j^z= |t^+\rag\lag t^+| -|t^-\rag\lag t^-|$ and $\tilde{S}_j^\pm= \sqrt{2} \lt(|t^\pm\rag\lag t^0| + |t^0\rag\lag t^\mp|\rt)$, by projecting spin operators onto the triplet state through
\begin{align}
&\mcl{P}_j (\bm{S}_{1,j}+\bm{S}_{3,j}) \mcl{P}_j=\mcl{P}_j (\bm{S}_{2,j}+\bm{S}_{4,j}) \mcl{P}_j \notag\\
&\hspc =\frac{1}{2}\mcl{P}_j \lt(\sum_i\bm{S}_{i,j}\rt) \mcl{P}_j=\frac{1}{2}\tilde{\bm{S}}_j,
\end{align}
with the projection operator $\mcl{P}_j=\sum_\gamma |t^\gamma\rag_j\lag t^\gamma|_j$.
Consequently, the inter cluster interactions with $\phi=\frac{\pi}{4}$ give the Heisenberg-type interactions between neighboring effective spins [condition (III)],
\begin{equation}
\tilde{\mcl{H}}=\lt(\prod_j\mcl{P}_j\rt)\lt(\sum_j\mcl{H}_\inter^{(j)}\rt)\lt(\prod_j\mcl{P}_j\rt)=\frac{J_\inter}{8}\sum_j\tilde{\bm{S}}_j\cdot\tilde{\bm{S}}_{j+1}.
\end{equation}
Assuming $J_\inter \ll J_\intra$ [condition (II)], we should obtain CBHS as the ground state in the 4-spin cluster chain.
To confirm the existence of the CBHS, we have calculated the magnetization curve and correlation functions using the variational matrix product state (VMPS) method [Fig.~3(c)]~\cite{Schollwock11,Note2}.
We can see the magnetization plateau at magnetization $M=1$ [see the inset of Fig.~3(c)] because of the free spins at the edges of chain, which indicates the existence of the CBHS.
In Fig.~3(d), we can find another evidence of the CBHS that the string correlation converges on a constant value, while the spin correlation decreases exponentially with increasing the cluster-cluster distance.

As compared with the first example, the $N=\>$odd cases require a little more burden to find the CBHS due to the Kramers' theorem: the energy levels of the system containing an odd number of electrons/spins in the presence of the time reversal symmetry should have even-fold degeneracy.  
Obviously, this theorem interferes with the condition (I).
To overcome this inconvenience, we introduce the magnetic field into the model Hamiltonian $\mcl{H}_0$ (1), which breaks the time reversal symmetry and recover the condition (I) to obtain the CBHS in an $N=\>$odd SCC.
Since the magnetic field is easily controlled in the laboratory, this procedure is not against to our purpose for designing the Haldane state with one-half spins for the practical use.
The magnetic field is introduced by the Zeeman term: $\mcl{H}=\mcl{H}_0-h^z\sum_{i,j} S_{i,j}^z$.
Since the magnetization $M=\sum_{i,j} S_{i,j}^z$ is preserved in the Hamiltonian, the magnetic field just adds the Zeeman energy $-h^z M$ to the zero-field energy.
Therefore, if the following conditions are satisfied at zero field instead of condition (I), a field-induced triplet (FIT) emerges as the ground state at the critical field $h_c^z$.
\begin{enumerate}
\item[i.] The lowest energy $\epsilon_S$ for the total spin $S=S_0,S_0+1,S_0+2$ satisfies the relation that $\Delta=\epsilon_{S_0+2}-\epsilon_{S_0+1}=\epsilon_{S_0+1}-\epsilon_{S_0}>0$. Here the critical field is given by $h_c^z=\Delta$ [see Fig.~4(a)].
\item[ii.] The lowest energy $\epsilon_S$ for the total spin $S_0$ gives the ground-state energy together with the Zeeman energy at $h^z=h_c^z$: $\epsilon_{S_0}-h_c^zS_0\leq \epsilon_{S^\prime}-h_c^zS^\prime$ for arbitrary total spin $S^\prime$.
\end{enumerate}

\begin{figure}
\centering
\includegraphics[width=0.5\textwidth]{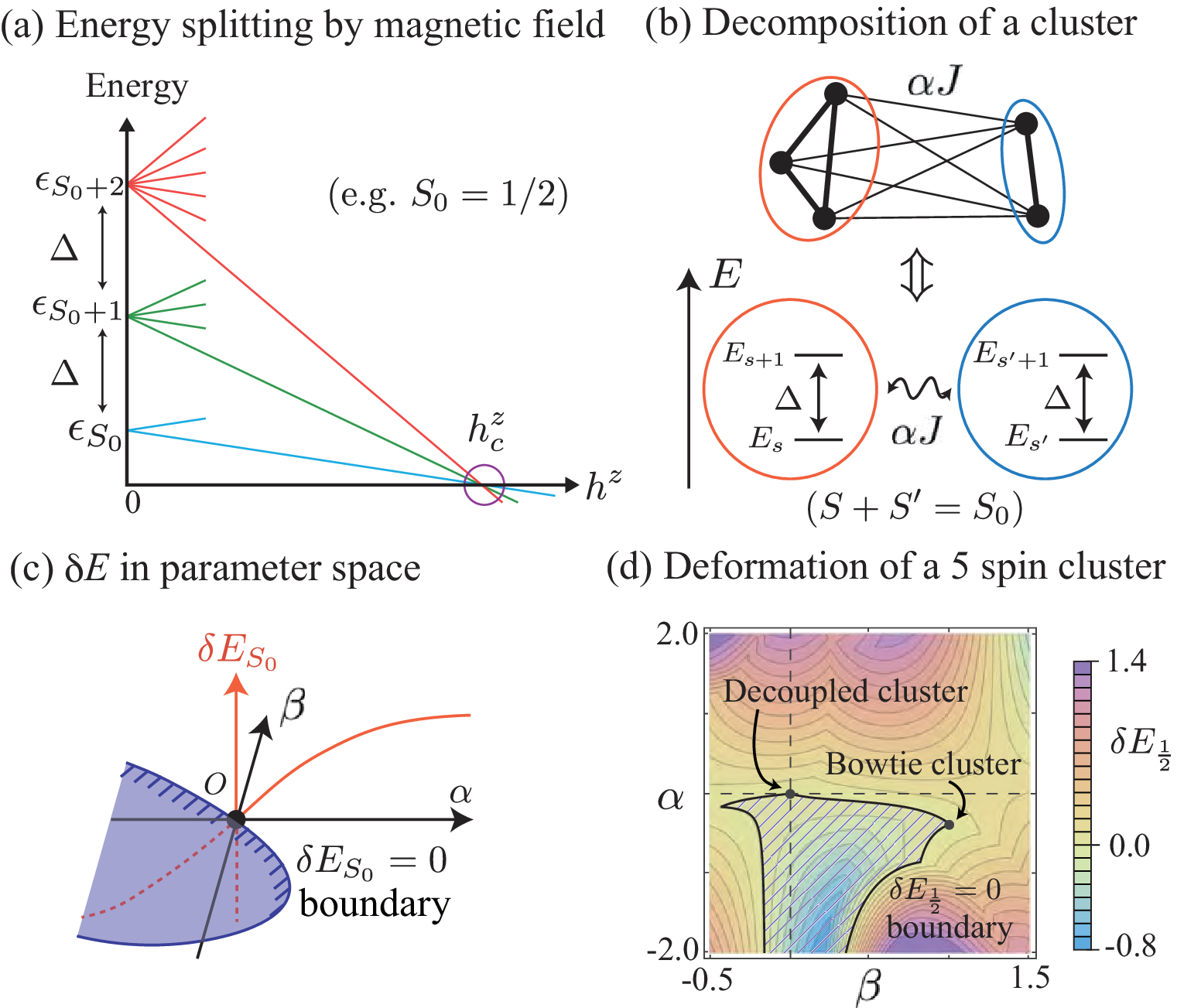}
\caption{(a) Schematic of condition (i). If $\epsilon_{S_0+2}-\epsilon_{S_0+1}=\epsilon_{S_0+1}-\epsilon_{S_0}=\Delta$ without magnetic fields, a field-induced triplet (FIT) consisting of $|S^z\rag=|\frac{1}{2}\rag$, $|\frac{3}{2}\rag$, and $|\frac{5}{2}\rag$ for $S_0=1/2$, emerges at the critical field $h_c^z$. (b) Weakly coupled two subsystems denoted by orange and blue circles in a 5-spin cluster. The two subsystems are connected by an exchange interaction $\alpha J$ ($J>0$ is the energy unit of cluster). (c) $\delta E_{S_0}=\epsilon_{S_0+2}-2\epsilon_{S_0+1}+\epsilon_{S_0}$ as a function of $J$ and a deformation parameter $\beta$ independent of $\alpha$. The $\beta=0$ case corresponds to (b). The white (purple) region denotes the (positive) negative $\delta E_{S_0}$. Since $\delta E_{S_0}$ changes its sign at $\alpha=0$ without deformation ($\beta=0$), we can deform the model parameters along the $\delta E_{S_0}=0$ line, on which the condition (i) is satisfied. (d) An example of deformation of a 5 spin cluster. The point of $(\alpha,\beta)=(0,0)$ corresponds to the trivial case represented in (b). We can reach a simpler model, a {\it bowtie} cluster, in Fig.~5(a), as we trace the boundary.}
\label{fig4}
\end{figure}

How do we find solutions of the condition (i) for the $N=\>$odd spin cluster?
The condition (i) requires three different total spins $S=S_0,S_0+1,S_0+2$, so that $N\geq5$ is required.
The number of bonds in an $N\geq5$ cluster is too large to search solutions in the full parameter space, e.g., a 5-spin cluster has 10 bonds.
In other words, we have to search the solution in ten-dimensional parameter space.
Instead, we propose the following procedure in the present study.
Firstly, we find a trivial solution in a cluster where two subsystems having the same gap $\Delta$ are weakly coupled via an exchange energy $\alpha J$ ($J>0$ is the energy unit of cluster and $\alpha \ll \Delta/J$) as shown in Fig.~4(b).
In this model, $\alpha=0$ corresponds to a trivial solution of the condition (i), i.e., $\delta E_{S_0}=\epsilon_{S_0+2}-2\epsilon_{S_0+1}+\epsilon_{S_0}=0$, and the sign of $\delta E_{S_0}$ changes from negative to positive at $\alpha=0$ with increasing $\alpha$ as explained below.
Next, we introduce a parameter $\beta$ independent from $\alpha$ in order to deform the model parameters.
The parameter $\beta$ is chosen as one of model parameters to deform the model into a simpler one: $\beta=0$ ($\beta=1$) corresponds to the trivial model (a simpler model).
For instance, $\beta$ is introduced to decrease the number of interactions or to equalize the strength of several interactions.
In the parameter plane of $\alpha$ vs. $\beta$, the trivial point is located at $(\alpha,\beta)=(0,0)$.
Since $\delta E_{S_0}$ is continuous in the parameter space, the boundary between the regions of positive and negative $\delta E_{S_0}$ spreads around the trivial point $(\alpha,\beta)=(0,0)$ [see Fig.~4(c)].
Hence, we can trace the boundary satisfying the condition (i), i.e., $\delta E_{S_0}=0$, by deforming model parameters from a trivial point to a non-trivial one correponding to a simpler model in the parameter space.

As mentioned above, we firstly start from giving a trivial model of a cluster with odd-number spins, where two subsystems of the cluster weakly couples via an exchange energy $\alpha J$.
Figure 4(b) shows a 5-spin cluster as an example of $N=\>$odd clusters, where the cluster is devided into two subsystems of three and two spins.
If the two subsystems have the same energy gap $\Delta$ between the lowest levels of the total spin $S$ ($S^\prime$) and $S+1$ ($S^\prime+1$), the lowest energy of the total system for the total spin $S_0$, $S_0+1$, and $S_0+2$ with $S_0=S+S^\prime$ is given by,
\begin{align}
\epsilon_{S_0}&=\epsilon_S+\epsilon_{S^\prime}+\alpha JSS^\prime, \\ 
\epsilon_{S_0+1}&=\begin{cases}
\epsilon_S+\epsilon_{S^\prime}+\Delta+\alpha J(S+1)S^\prime& (J\geq 0)\\
\epsilon_S+\epsilon_{S^\prime}+\Delta+\alpha JS(S^\prime+1) & (J<0)  
\end{cases},\\
\epsilon_{S_0+2}&=\epsilon_S+\epsilon_{S^\prime}+2\Delta+\alpha J(S+1)(S^\prime+1).
\end{align}
Since $\delta E_{S_0}$ reads
\begin{equation}
\delta E_{S_0}=\begin{cases}
\alpha J(S-S^\prime+1) & (\alpha\geq 0),\\
\alpha J(S^\prime-S+1) & (\alpha<0),
\end{cases}
\end{equation}
the condition (i) given by $\delta E_{S_0}=0$ is satisfied when $\alpha=0$.
We note that $\delta E_{S_0}$ changes its sign at $\alpha=0$ if $S=S^\prime+\frac{1}{2}$.
This sign change is useful for finding a simpler model in the parameter space of $\alpha$ and $\beta$.

Figure 4(c) shows a schematic behavior of $\delta E_{S_0}$ in the $\alpha$ vs. $\beta$ space.
The white (purple) region in the $\alpha$-$\beta$ plane denotes the (positive) negative $\delta E_{S_0}$, so that the boundary of the two regions denoted by the solid black line stretches around the trivial point $(\alpha,\beta)=(0,0)$.
Consequently, we can deform the model parameters by tracing the $\delta E_{S_0}=0$ line from the trivial point [see Fig.~4(c)].
Based on this strategy, we have tried several deformations of parameters by changing $\beta$ on the boundary to obtain a simpler model.
As an example, we can find the following {deformed Hamiltonian of a 5-spin cluster with an additional angle $\lambda$ controlling interactions in each subsystem:
\begin{equation}
\mcl{H}(\lambda,\alpha,\beta)^{\mrm{(clst)}}=J\lt[\mcl{H}_1^{\mrm{(clst)}}+\mcl{H}_2^{\mrm{(clst)}}+\mcl{H}_3^{\mrm{(clst)}}\rt]
\end{equation}
with
\begin{align}
&\mcl{H}_1^{\mrm{(clst)}} = \cos\lambda \bm{S}_{1}\cdot\bm{S}_{2} +\frac{\sin\lambda}{\sqrt{2}} (\bm{S}_{1}+\bm{S}_{2})\cdot\bm{S}_{3},\\
&\mcl{H}_2^{\mrm{(clst)}} = \xi_\beta \bm{S}_{4}\cdot\bm{S}_{5},\\ 
&\mcl{H}_3^{\mrm{(clst)}} = \alpha \Big[\frac{\sin\zeta_\beta}{2} (\bm{S}_{1}+\bm{S}_{2})\cdot(\bm{S}_{4}+\bm{S}_{5})\notag \\
& \hspace{5em} +\frac{\cos\zeta_\beta}{\sqrt{2}}\bm{S}_{3}\cdot(\bm{S}_{4}+\bm{S}_{5})]\Big],
\end{align}
where the deformation parameter $\beta$ determines two coefficients
\begin{align}
\xi_\beta=\cos\lambda+(1-\beta) \frac{\sin\lambda}{2\sqrt{2}}, \hspc \zeta_\beta=(1-\beta) \tan^{-1}(\sqrt{2})+\beta \pi.
\end{align}
In this model, $\beta=0$ gives the situation of Fig.~4(b), i.e., $\mcl{H}_1^\mrm{(clst)}$ for one subsystem $\{\bm{S}_1,\bm{S}_2,\bm{S}_3\}$and $\mcl{H}_2^\mrm{(clst)}$ for the other $\{\bm{S}_4,\bm{S}_5\}$, which are connected by the weak coupling $\alpha$, have the same energy gap.
On the other hand, $\beta=1$ corresponds to a simpler model where the coupling constant $\sin\zeta_\beta=0$, $\cos\zeta_\beta=-1$, and $\xi_\beta=\cos\lambda$.
Figure 4(d) shows the color map of $\delta E_{\frac{1}{2}}$ in the $\alpha$-$\beta$ plane with a fixed $\lambda=\frac{\pi}{8}$.
We can find that the $\delta E_{S_0}=0$ boundary connects between $(\alpha,\beta)=(0,0)$ and $(\alpha,\beta)=(\alpha_c,1)$ with $\alpha_c=-\sin\lambda$.
We thus obtain the following simpler model, which we call a {\it bowtie} cluster, as a solution of the condition (i).
\begin{align}
\mcl{H}(\lambda,\alpha_c,1)^{\mrm{(clst)}}&=\cos\lambda (\bm{S}_{1}\cdot\bm{S}_{2}+\bm{S}_{4}\cdot\bm{S}_{5})\notag \\
&\hspc +\frac{\sin\lambda}{\sqrt{2}} \bm{S}_3\cdot(\bm{S}_{1}+\bm{S}_{2}+\bm{S}_{4}+\bm{S}_{5}).
\end{align}

{\it Second example: a 5-spin cluster chain (an $N=5$ SCC).} ---
As discussed above, we have found that the 5 spin cluster described by $\mcl{H}(\lambda,\alpha_c,1)^{\mrm{(clst)}}$ (17) satisfies the condition (i) for $\lambda=\frac{\pi}{8}$.
To confirm the CBHS in the SCC constructed by $\mcl{H}(\lambda,\alpha_c,1)^{\mrm{(clst)}}$, we consider the following SCC model [see Fig.~5(a)].
\begin{align}
\mcl{H}_{\mrm{intra}}^{(j)} &= \frac{J_\intra}{\sqrt{2}} \Big\{\cos\lambda\lt[[\mrm{Hs}^{(j)}_{(1,2)}+\mrm{Hs}^{(j)}_{(4,5)}\rt] \notag\\
& \htab +\frac{\sin\lambda}{\sqrt{2}}\lt[[\mrm{Hs}^{(j)}_{(1,3)}+\mrm{Hs}^{(j)}_{(2,3)}+\mrm{Hs}^{(j)}_{(3,4)}+\mrm{Hs}^{(j)}_{(3,5)}\rt]\Big\}
\end{align}
and 
\begin{align}
\mcl{H}_{\mrm{inter}}^{(j)} &= \frac{J_\inter}{2} \Big[\cos\eta({\mrm{Hs}^\prime}^{(j,j+1)}_{(2,1)}+{\mrm{Hs}^\prime}^{(j,j+1)}_{(5,4)}) \notag\\
& \htab +\sin\eta({\mrm{Hs}^\prime}^{(j,j+1)}_{(2,4)}+{\mrm{Hs}^\prime}^{(j,j+1)}_{(5,1)})\Big],
\end{align}
Figure 5(b) shows the lowest energy for the total spin $S=\frac{1}{2},\frac{3}{2},\frac{5}{2}$ and $\delta E_{\frac{1}{2}}$.
Interestingly, we can see that the condition (i) for $S_0=\frac{1}{2}$ is satisfied in $\lambda \in [0,\lambda_1]$ or $[\lambda_2,2\pi]$~\cite{Note3}. 
When $\lambda \in[\lambda_2,2\pi]$, $\epsilon_{\frac{1}{2}}$ gives the lowest energy, which is the same as the case for $\lambda \in[0,\lambda_1]$.

We therefore expect the FIT as the ground state of a cluster at the critical magnetic field [condition (ii)], where the effective spin operator with $S=1$ is given by $\tilde{S}_j^z=|\frac{5}{2}\rag_j\lag\frac{5}{2}|_j-|\frac{1}{2}\rag_j\lag\frac{1}{2}|_j$.
Since the projection operator into the FIT space is given by $\mcl{P}_j=\sum_{m=\frac{1}{2},\frac{3}{2},\frac{5}{2}}|m\rag_j\lag m|_j$, we obtain the projected magnetization of a cluster $\mcl{P}_jM_j\mcl{P}_j=\sum_{i} \mcl{P}_jS_{i,j}^z\mcl{P}_j=\sum_{m=\frac{1}{2},\frac{3}{2},\frac{5}{2}}m\,|m\rag_j\lag m|_j$, so that the $z$ component of effective spin corresponds to the magnetization of a cluster decreased by $\frac{3}{2}$, $\tilde{S}_j^z=\mcl{P}_j\lt(M_j-\frac{3}{2}\rt)\mcl{P}_j$.
This correspondence means that the effective $S=1$ spin states are given by $|\tilde{1}\rag_j\cong |\frac{5}{2}\rag_j$,$|\tilde{0}\rag_j\cong |\frac{3}{2}\rag_j$, and $|\tilde{-1}\rag_j\cong |\frac{1}{2}\rag_j$.
Thus, the Haldane gap can emerge at $\lag M_{j}\rag =\frac{3}{2}$, i.e., $M=\sum_j\lag M_{j}\rag=\frac{3}{5}M_{\mrm{sat}}$ where $M_{\mrm{sat}}=\frac{5}{2}L$ is the saturated magnetization.
Note that the CBHS in an $N=\>$odd SCC is induced by the magnetic field, so that we call it {\it field-induced} Haldane state.
To confirm this gap, we have calculated magnetization curve with a weak inter cluster interaction using the VMPS method [see Fig.~5(c)].
In Fig.~5(c), the small magnetization plateau is observed at $M=\frac{3}{5}M_{\mrm{sat}}+1$, because the Haldane state has free spins at the edges of chain.
Moreover, we confirm the existence of string order based on the effective spins in Fig.~5(d), while the spin correlation decreases exponentially.

\begin{figure*}
\centering
\includegraphics[width=0.9\textwidth]{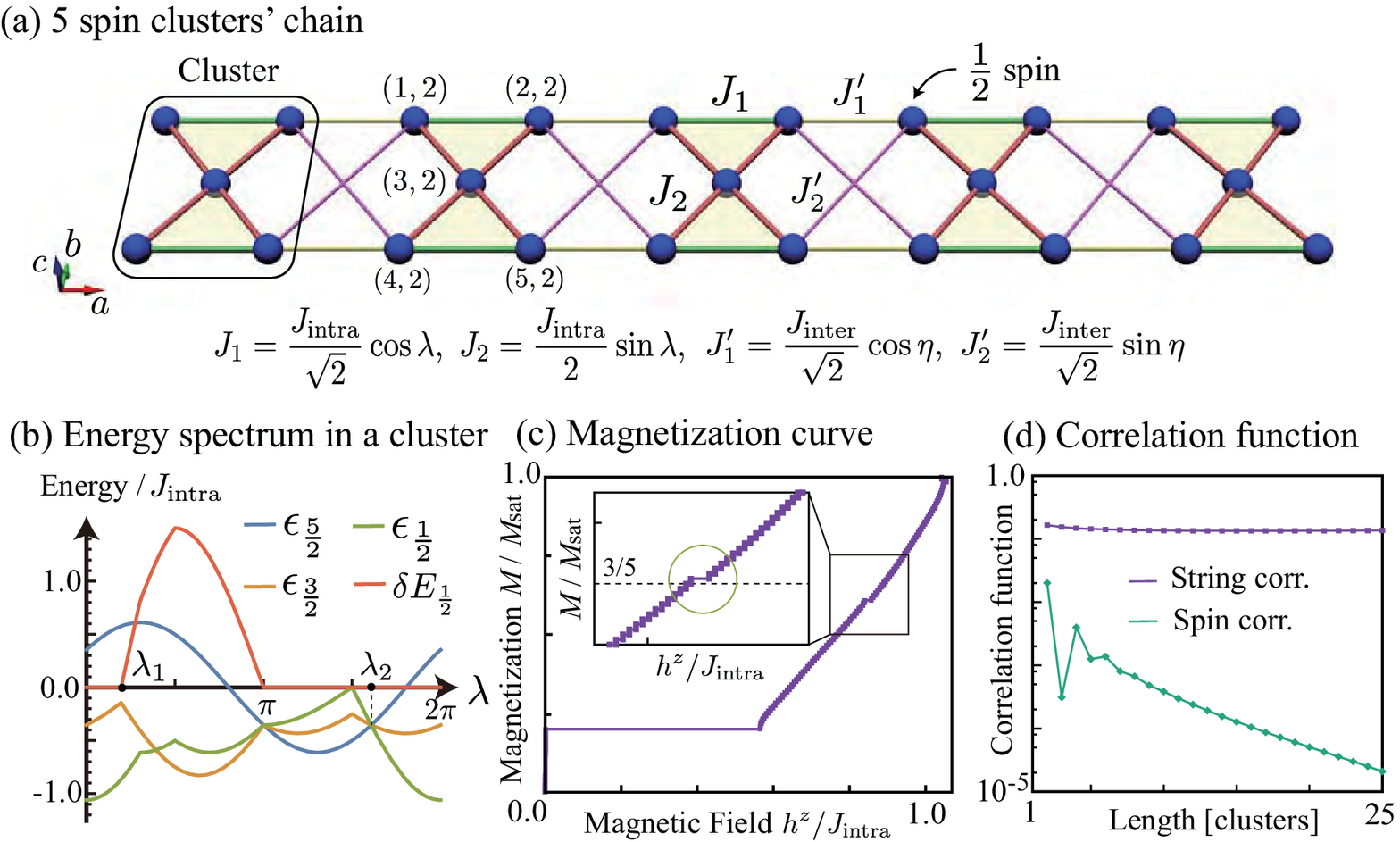}
\caption{(a) Five-spin cluster chain (an $N=5$ SCC). Notations are the same as Fig.~2(a). (b) Energy of a cluster as a function of $\lambda$ that controls $J_i$ with fixed $\eta=\pi/4$. Here $\lambda_1=\frac{\pi}{4}-\frac{1}{2}\sin^{-1}\lt(\frac{1}{3}\rt)$ and $\lambda_2=\frac{3}{2}\pi+\sin^{-1}\lt(\frac{1}{3}\rt)$. (c) Magnetization curve normalized by $M_{\mrm{sat}}$ and (d) absolute value for string and spin correlation functions for $J_\inter/J_\intra=0.3$, $\lambda=\frac{\pi}{8}$, and $\eta=\frac{\pi}{4}$ in a 48-cluster chain ($L=48$) with open boundary condition. The inset of (c) is an enlarged view around $M=\frac{3}{5}M_{\mrm{sat}}$, where we find the magnetization plateau at $M=\frac{3}{5}M_{\mrm{sat}}+1$ denoted by a circle. The correlation functions in (d) are calculated at $M=\frac{3}{5}M_{\mrm{sat}}+1$.}
\label{fig5}
\end{figure*}

In summary, we have studied the CBHS emerging in SCCs at low temperatures, motivated by the discovery of the CBHS in Fedotovite.
The CBHSs have been found in SCCs of one-half spins which contain not only an even number of spins but also an odd number of spins.
For the cluster with the even-number spins, the three conditions (I), (II), and (III) are useful for finding the CBHS in SCCs.
For the odd number case, the condition (I) is replaced by the two conditions (i) and (ii) and magnetic fields have to be applied.
To find a simpler model that exhibits the Haldane state for the odd number case, we have proposed a procedure of deforming parameters starting from a trivial point where the condition (i) is satisfied in the original parameter space.
Using this procedure, we have found a simpler model exhibiting the CBHS evidenced by the magnetization plateau and the string order, i.e., a field-induced Haldane state.
Since the Haldane state is proposed to be a possible holographic quantum computer, our concept of designable spin systems exhibiting the CBHS is useful for its application.

\begin{acknowledgments}
We would like to thank M. Fujihala and S. Mitsuda for valuable discussions. 
This work was partly supported by Grant-in-Aid for Young Scientists (B) (Grant No.16K17753), the CDMSI project on a post-K computer, and the inter-university cooperative research program of IMR, Tohoku University.
Numerical computation in this work was carried out on the supercomputers at JAEA and the Supercomputer Center at Institute for Solid State Physics, University of Tokyo.
\end{acknowledgments}


\begin{thebibliography}{99}\label{sec:TeXbooks}
\bibitem{Read00} N. Read and D. Green, Phys. Rev. B {\bf 61}, 10267 (2000).
\bibitem{Kitaev01} A. Y. Kitaev: Phys.-Usp. {\bf 44}, 131 (2001).
\bibitem{Ivanov01} D. A. Ivanov, Phys. Rev. Lett. {\bf 86}, 268 (2001).
\bibitem{Mazza13} L. Mazza, M. Rizzi, M. D. Lukin, and J. I. Cirac, Phys. Rev. B {\bf 88}, 205142 (2013).

\bibitem{Tang11} E. Tang, J.-W. Mei, and X.-G. Wen, Phys. Rev. Lett. {\bf 106}, 236802 (2011).
\bibitem{Sun11} K. Sun, Z. Gu, H. Katsura, and S. D. Sarma, Phys. Rev. Lett. {\bf 106}, 236803 (2011).
\bibitem{Neupert11} T. Neupert, L. Santos, C. Chamon, and C. Mudry, Phys. Rev. Lett. {\bf 106}, 236804 (2011).

\bibitem{Kopnin11} N. B. Kopnin, T. T. Heikkil\"a, and G. E. Volovik, Phys. Rev. B {\bf 83}, 220503(R) (2011).

\bibitem{Haldane83} F. D. M. Haldane, Phys. Lett. {\bf 93A}, 464 (1983); Phys. Rev. Lett. {\bf 50}, 1153 (1983).

\bibitem{Note0} In this article, we use the natural unit $\hbar=1$.

\bibitem{Affleck88} I. Affleck I, T. Kennedy, E. Lieb, and H. Tasaki, Commun. Math. Phys. {\bf 115}, 477 (1988).
\bibitem{Nijs89} M. den Nijs and K. Rommelse, Phys Rev. B {\bf 40}, 4709 (1989).
\bibitem{Tasaki91} H. Tasaki, Phys. Rev. Lett. {\bf 66}, 798 (1991).
\bibitem{Kennedy92} T. Kennedy and H. Tasaki, Phys. Rev. B {\bf 45} 304 (1992).

\bibitem{Pollmann12} F. Pollmann, E. Berg, A. M. Turner, and M. Oshikawa, Phys. Rev. B {\bf 85}, 075125 (2012).
\bibitem{Miyake08} G. K. Brennen and A. Miyake, Phys. Rev. Lett. {\bf 101}, 010502 (2008); A. Miyake, Phys. Rev. Lett. {\bf 105}, 040501 (2010).

\bibitem{Gross07} D. Gross and  J. Eisert, Phys. Rev. Lett. {\bf 98}, 220503 (2007).
\bibitem{Else12} D. V. Else, I. Schwarz, S. D. Bartlett, and A. C. Doherty, Phys. Rev. Lett. {\bf 108}, 240505 (2012).

\bibitem{Ishiguro98} As a review, see T. Ishiguro, K. Yamaji, and G. Saito, Organic Superconductors (Springer-Verlag, Berlin, 1998).

\bibitem{Kane98} B. E. Kane, Nature {\bf 393}, 133 (1998).
\bibitem{Loss98} D. Loss and D. P. DiVincenzo, Phys. Rev. A {\bf 57}, 120 (1998).

\bibitem{Cirac95} J. I. Cirac and P. Zoller, Phys. Rev. Lett. {\bf 74}, 4091 (1995).
\bibitem{Greiner02} M. Greiner, O. Mandel, T. Esslinger, T. W. H\"ansch, and I. Bloch, Nature {\bf 415}, 39 (2002).

\bibitem{Vergasova88} L. P. Vergasova, S. K. Filatov, Y. K. Serafimova, and G. L. Starova, Doklady Acad. Nauk SSSR {\bf 299}, 961 (1988).
\bibitem{Starova91} G. L. Starova, S. K. Filatov, V. S. Fundamensky, and L. P. Vergasova, Mineral. Mag. {\bf 55}, 613 (1991).
\bibitem{Fujihala18} M. Fujihala, T. Sugimoto, T. Tohyama, S. Mitsuda, R. A. Mole, D. H. Yu, S. Yano, Y. Inagaki, H. Morodomi, T. Kawae, H. Sagayama, R. Kumai, Y. Murakami, K. Tomiyasu, A. Matsuo, and K. Kindo, Phys. Rev. Lett. {\bf 120}, 077201 (2018).

\bibitem{Masuda06} T. Masuda, A. Zheludev, H. Manaka, L.-P. Regnault, J.-H. Chung, and Y. Qiu, Phys. Rev. Lett. {\bf 96}, 047210 (2006).
\bibitem{Vekua06} T. Vekua and A. Honecker, Phys. Rev. B {\bf 73}, 214427 (2006).

\bibitem{Note1} Note that these conditions are not the necessary conditions but sufficient ones.

\bibitem{Schollwock11} For example, see U. Schollw\"ock, Annal. Phys. {\bf 326}, 965 (2011).
\bibitem{Note2} In VMPS calculations, we set the number of kept states $m=300$, as we confirm that the error of truncated states is less than $10^{-8}$. 


\bibitem{Note3} The region $[\pi,\lambda_2]$ is not suitable for the consdition (i) because the gap $\Delta$ is negative, i.e., $\epsilon_{\frac{5}{2}}$ ($\epsilon_{\frac{3}{2}}$) is lower than $\epsilon_{\frac{3}{2}}$ ($\epsilon_{\frac{1}{2}}$).
\end{thebibliography}
\end{document}